\documentclass{nature}

\usepackage{mathptmx}
\usepackage[T1]{fontenc}

\usepackage{amsmath}
\usepackage{amssymb}
\usepackage{graphicx}
\usepackage{threeparttable}
\usepackage{bibunits}
\usepackage{etoolbox}
\newcounter{mybibitem}
\AtBeginEnvironment{bibunit}{\setcounter{mybibitem}{0}}
\makeatletter

\patchcmd{\thebibliography}{ \usecounter{enumiv}}{  \usecounter{enumiv}  \setcounter{enumiv}{\value{mybibitem}}   }{}{}

\title{Rotation in [C{\sc ii}]-emitting gas in two galaxies at a redshift of 6.8
}

\author{Renske Smit$^{1,2}$, Rychard J. Bouwens$^3$, Stefano Carniani$^{1,2}$, Pascal A. Oesch$^4$, Ivo Labb\'{e}$^3$, Garth D. Illingworth$^5$, Paul van der Werf$^3$, Larry D. Bradley$^6$, Valentino Gonzalez$^{7,8}$, Jacqueline A. Hodge$^3$,  Benne W. Holwerda$^9$, Roberto Maiolino$^{1,2}$, Wei Zheng$^{10}$
\vspace{5mm}}

\begin{document}\sloppy
\begin{bibunit}[naturemag]
\twocolumn

\maketitle

\begin{affiliations}
{\small
 \item Cavendish  Laboratory, University of Cambridge, 19 JJ Thomson Avenue, Cambridge CB3 0HE, UK
 \item Kavli Institute for Cosmology, University of Cambridge, Madingley Road, Cambridge CB3 0HA
 \item Leiden Observatory, Leiden University, NL-2300 RA Leiden, Netherlands 
  \item Geneva Observatory, University of Geneva, Ch. des Maillettes 51, CH-1290 Versoix, Switzerland
 \item UCO/Lick Observatory, University of California, Santa Cruz, 1156 High St, Santa Cruz, CA 95064, USA
 \item Space Telescope Science Institute, 3700 San Martin Drive Baltimore, MD 21218
 \item Departmento de Astronomia, Universidad de Chile, Casilla 36-D, Santiago, Chile
 \item Centro de Astrofisica y Tecnologias Afines (CATA), Camino del Observatorio 1515, Las Condes, Santiago, Chile
 \item Department of Physics and Astronomy, University of Louisville, Louisville KY 40292 USA
 \item  Department of Physics and Astronomy, The Johns Hopkins University, 3400 North Charles Street, Baltimore, MD 21218, USA

}
\end{affiliations}

\begin{abstract}
The earliest galaxies are expected to emerge in the first billion years of
the Universe during the Epoch of Reionization.   However, both the
spectroscopic confirmation of galaxies at this epoch and the 
characterization of their early dynamical state has been hindered by the lack
of bright, accessible lines to probe the velocity structure of their
interstellar medium.  We present the first spectroscopic confirmation of near-infrared selected sources at $z>6$
using the far-infrared [\ion{C}{2}]\,$\lambda$157.74$\micron$  emission line,  and, for the first
time, measurement of the velocity structure, for two galaxies at  $z=6.8540\pm0.0003$ and $z=6.8076\pm0.0002$.
 Remarkably, the
 [\ion{C}{2}] line luminosity from these galaxies is higher
than previously found in `normal' star-forming galaxies at $z>6.5$\cite{Ouchi2013,Ota2014,Maiolino2015,Knudsen2016,Pentericci2016}.
This suggests that we are sampling a part of the galaxy population different from the galaxies found through detection of the Ly$\alpha$ line.
The luminous and extended  [\ion{C}{2}] detections
reveal clear velocity gradients that, if interpreted as rotation, would suggest these galaxies have similar dynamical properties as the turbulent, yet rotation-dominated disks that have been observed for H$\alpha$ emitting galaxies 2 Gyr later at cosmic noon. Our novel approach for
confirming galaxies during Reionization paves the way for larger studies of
distant galaxies with spectroscopic redshifts. Particularly
important, this opens up opportunities for high
angular-resolution [\ion{C}{2}] dynamics in galaxies less than one billion years
after the Big Bang.

\end{abstract}

\begin{figure*}[h]
\centering
\includegraphics[width=0.99\textwidth,trim=0mm 0mm 0mm -2mm] {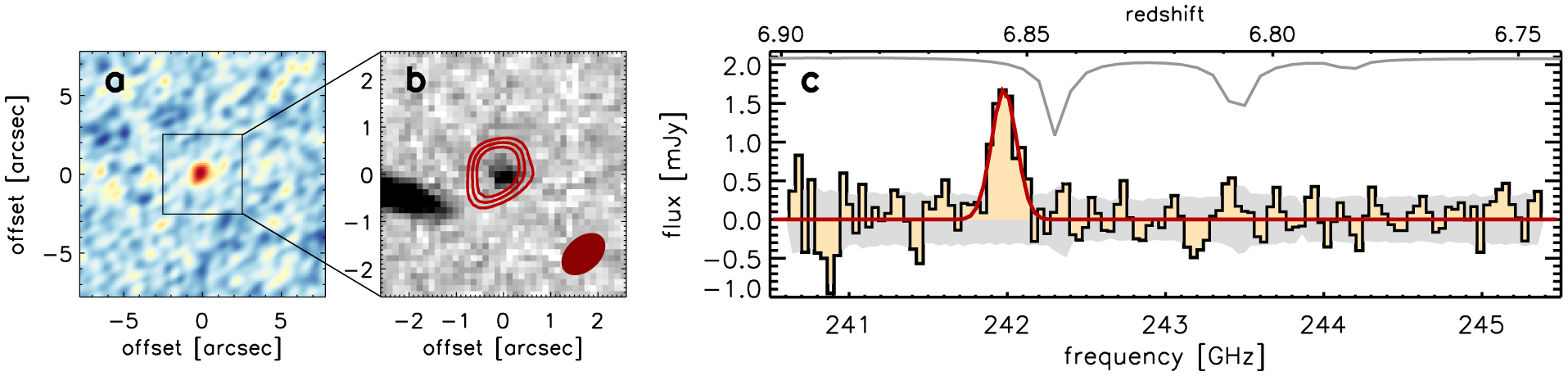}  
\includegraphics[width=0.99\textwidth,trim=0mm 0mm 0mm -2mm] {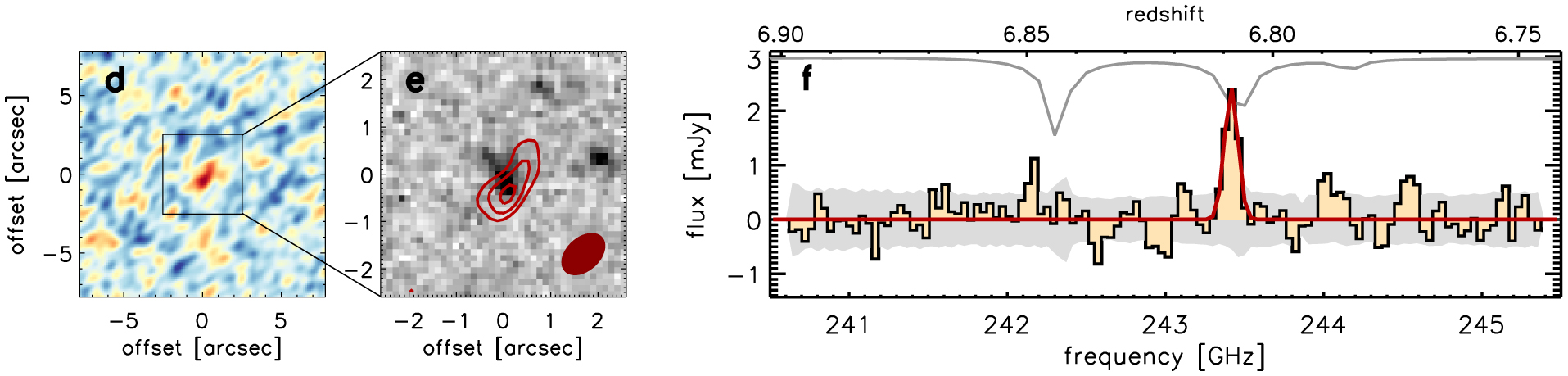} 
\caption{\textbf{| Spectroscopic line confirmations of the targeted galaxies in this study.} ALMA line maps and spectra of two galaxies with photometric redshifts in the range $6.6<z_{\rm phot}<6.9$ \cite{Smit2015}. We detect a $8.2\sigma$  [\ion{C}{2}] line at $z_{\rm [CII]}=6.8540\pm0.0003$ in galaxy COS-3018555981 (\textit{top panels}) and a $\sim5.1\sigma$  [\ion{C}{2}] line at $z_{\rm [CII]}=6.8076\pm0.0002$ in galaxy COS-2987030247 (\textit{bottom panels}). 
\textit{Left panels:} 20'' x 20'' images of the ALMA cube (before primary-beam correction) collapsed over 241.85-242.10 GHz and 243.35-243.45 GHz for COS-3018555981 and COS-2987030247 respectively (rms of 0.1 and 0.2 mJy respectively). 
\textit{Middle panels:} 5'' x 5'' zoom-in on the targeted sources. The $HST$ $H_{160}$ imaging is shown in grey-scale, while the overlaid red contours show the 3,4,5-$\sigma$ levels of the spectral line-averaged maps in the left panels. The filled ellipse in the bottom right corner indicates the beam size (1.1"$\times$0.7" half-power widths).
\textit{Right panels:} The spectra extracted within a contour of the half-maximum power in the line maps. The red line shows the best fit Gaussian line profile. The grey line at the top of the panels shows the atmospheric absorption, while the grey line at the bottom of the panels gives the measured rms for the spectrum (for clarity shown at a fixed offset of -1.0 and -1.5 mJy for the top and bottom panel respectively). 
}
\label{fig:COS3}
\end{figure*}

\begin{figure*}[h]
\centering
\includegraphics[width=.9\columnwidth,trim=30mm 163mm 95mm 30mm] {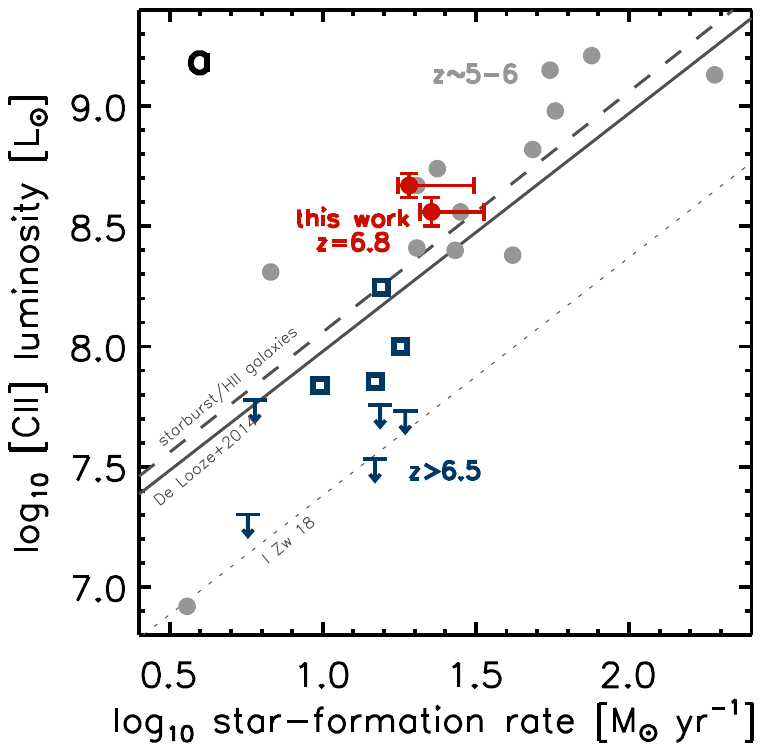} 
\includegraphics[width=.9\columnwidth,trim=30mm 163mm 95mm 30mm] {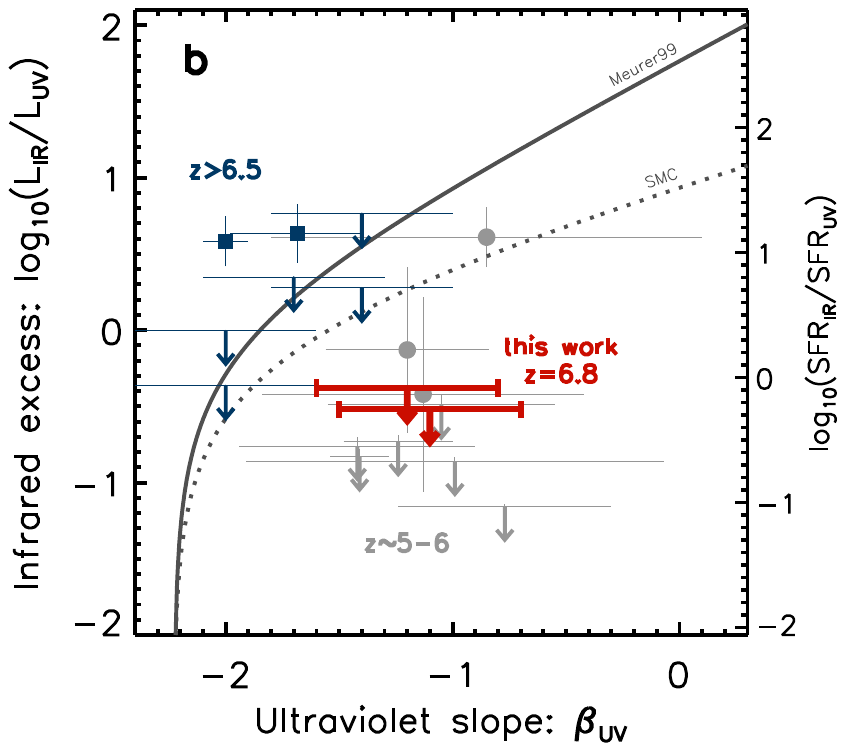} 
\caption{ \textbf{| [\ion{C}{2}] luminosity and dust continuum of $z>5$ galaxies.} \textit{Left panel:} The [\ion{C}{2}] line luminosity as a function of the star-formation rate, derived from the ultraviolet luminosity of COS-3018555981 and COS-2987030247 (\textit{red points}; uncertainty on the star-formation rate reflects the upper limits on the infrared continuum). 
We include [\ion{C}{2}] detections at redshift $z\sim5-6$ as light grey points \cite{Capak2015,Willott2015,Knudsen2016} and detections and upper limits at $z>6.5$ with blue open squares and arrows\cite{Pentericci2016,Maiolino2015,Ouchi2013,Ota2014}. 
The locally observed De Looze relations\cite{deLooze2014} are indicated by the solid (local star-forming galaxies) and dashed (starburst and HII galaxies) lines. Low-metallicity dwarfs are found to be systematically offset from the relation of local star-forming galaxies, with the local ultra low-metallicity dwarf galaxy I Zwicky 18 being offset by as much as 0.6 dex (dotted line).
\textit{Right panel:} The infrared excess (IRX=$L_{\rm UV}/L_{\rm IR}$) as a function of the UV-continuum slope ($\beta$) of our sources compared to the Meurer\cite{Meurer1999} relation (solid grey line) and a similar relation based on the dust law of the Small Magellanic Cloud\cite{Prevot1984} (dotted grey line).  
We include [\ion{C}{2}] detections at redshift $z\sim5-6$  as light grey points\cite{Capak2015} and detections and upper limits at $z>6.5$ with blue solid squares and arrows\cite{Watson2015,Schaerer2015,Laporte2017}. All upper limits are 1$\sigma$.
} 
\label{fig:SFRCII}
\end{figure*}

We have obtained spectroscopy with the Atacama Large Millimetre Array (ALMA) at 241--245\,GHz of two Lyman-break galaxies (LBGs) COS-3018555981 and COS-2987030247 at an estimated photometric redshift just below 7, roughly 800 million years after the Big Bang. These two sources are luminous in the rest-frame ultraviolet (UV; $L_{\rm UV}\sim2\times L^\ast$\cite{Stark2016}), but are still representative of `normal' star-forming galaxies at $z\sim7$ with a UV star-formation rate (SFR) of $19-23\,M_\odot\rm\,yr^{-1}$.  They were selected based on their blue rest-frame optical colours measured in the 3.6 and 4.5$\micron$ \textit{Spitzer} photometric bands\cite{Smit2015}, which strongly constrains the photometric redshift probability distribution to the redshift range $6.6<z<6.9$. 
These sources are among the most extreme [\ion{O}{3}]+H$\beta$ emitters known at $z\sim7$\cite{Smit2014,Smit2015}. We observed these sources in a 36 antennae configuration with ALMA (angular resolution of 1.1"$\times$0.7", equivalent to 5.8$\times$3.7 kpc at $z=6.8$) and 24 min on source integration time for each of the targets. We use this spectral scan to search for [\ion{C}{2}] in the redshift range $z_{\rm [CII]}=6.74-6.90$. Our results are summarized in Table \ref{tab:lines}.

We detect a line in each of the targets at $241.97\pm0.01\,\rm GHz$ and $243.42\pm0.01\,\rm GHz$, for COS-3018555981 and COS-2987030247 respectively, at $>5\sigma$ significance in both the one-dimensional spectra and the spectral line-averaged maps (Fig. \ref{fig:COS3}). We derive spectroscopic redshifts of $z_{\rm [CII]}=6.8540\pm0.0003$ and $z_{\rm [CII]}=6.8076\pm0.0002$, in excellent agreement with the photometric redshift estimates of $6.76\pm0.07$ and $6.66\pm0.14$ for COS-3018555981 and COS-2987030247, and line-widths of $232\pm30$ and $124\pm18$ $\rm km\,s^{-1}$ respectively. While successful line-searches have confirmed far-infrared lines in submillimetre selected star-bursting galaxies at $z>6$\cite{Riechers2013,Strandet2017}, and a few tantalising  "blind"   (i.e. with no optical or near-infrared counterpart) [\ion{C}{2}] emitters have recently been detected at $\sim4\sigma$\cite{Aravena2016}, this is the first time that normal star-forming galaxies in this early epoch, selected at optical or near-infrared wavelengths, have been confidently spectroscopically confirmed with ALMA.

We furthermore obtain upper limits on the far-infrared (FIR) dust continuum emission from the ALMA data. We find $\rm SFR_{\rm IR}<16-19\,M_\odot\rm\,yr^{-1}$, consistent with `normal' star-forming galaxies in the local Universe\cite{Leroy2012} and ruling out the presence of a dusty starburst in these sources. 
 Fig. \ref{fig:SFRCII} shows that for the colour of the UV-continuum slope ($\beta_{\rm UV}\sim-1.2$) in these galaxies we would expect a higher dust content (IRX=$L_{\rm IR}/L_{\rm UV}$) in these galaxies if they were consistent with having the Meurer dust law, which is locally observed to apply for starburst galaxies\cite{Meurer1999}. 
 Scatter in the IRX-$\beta_{\rm UV}$ relation can be due to dust geometry effect, galaxy population age or the shape of the attenuation curve.
However, for blue galaxies ($\beta_{\rm UV}\lesssim-0.5$) that scatter below the Meurer\cite{Meurer1999} relation, such as seen in our galaxies, the most likely way to reproduce the low values of IRX is through a steeper attenuation curve, such as has been derived for the Small Magellanic Cloud\cite{Prevot1984} (consistent with our measurements within 3$\sigma$), in combination with a potential increase in dust temperature at higher redshift.

\begin{figure*}
\centering
\includegraphics[width=0.45\textwidth,trim=-2mm 0mm -3mm 0mm] {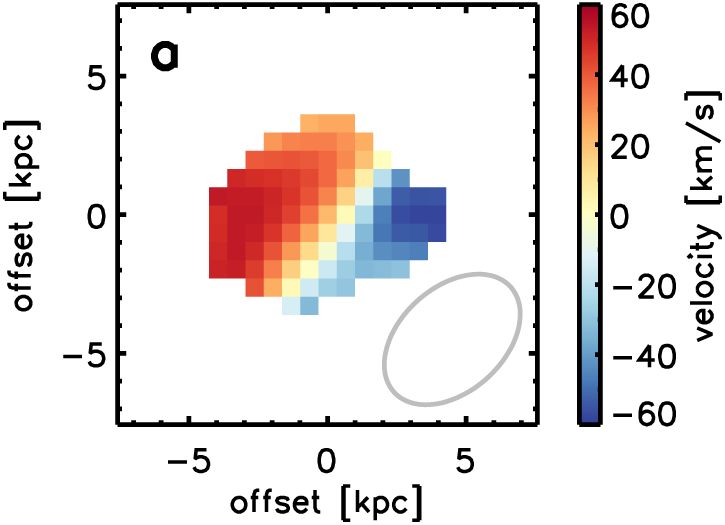} 
\includegraphics[width=0.45\textwidth,trim=-2mm 0mm -3mm 0mm] {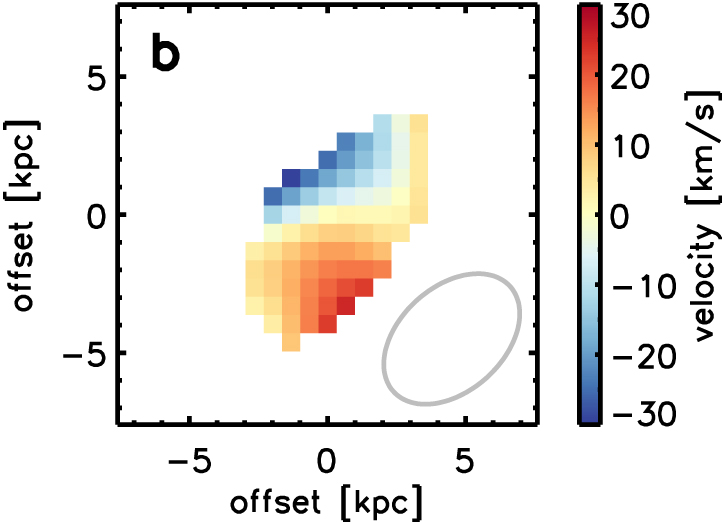} 
\caption{\textbf{| Velocity structure of the detected [\ion{C}{2}] in the two galaxies from this study.} The velocity field measured in our galaxies COS-3018555981 (\textit{left panel}) and COS-2987030247 (\textit{right panel}). The observations are spatially resolved, as shown by the beamsize of the observations indicated by the grey ellipse in the bottom right corner and reveal a projected velocity difference over the galaxy of $111\pm28$ and $54\pm20$ $\rm km\,s^{-1}$ respectively. Given the low angular resolution of the observations, the detected velocity gradients can be interpreted as disk rotation or potentially a merger with two or more velocity components. 
}
\label{fig:kin}
\end{figure*}

\begin{figure*}
\centering
\includegraphics[width=0.45\textwidth,trim=30mm 165mm 100mm 30mm] {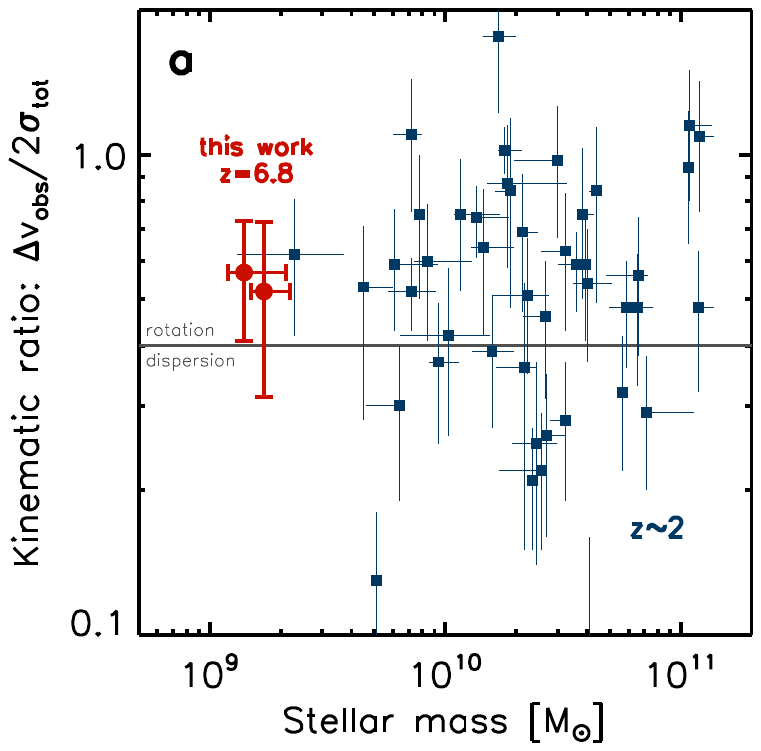} 
\includegraphics[width=0.45\textwidth,trim=30mm 165mm 100mm 30mm] {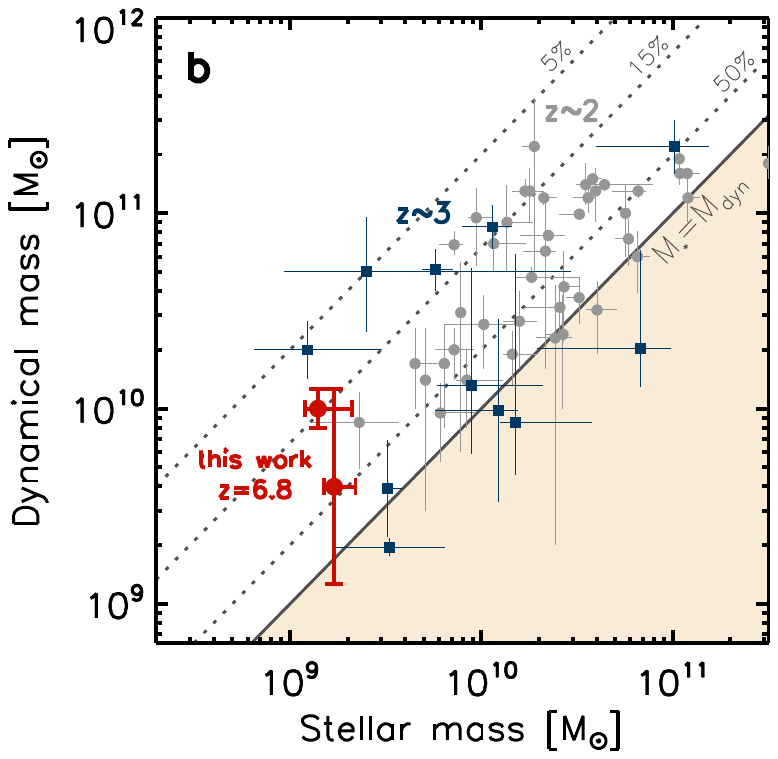} 
\caption{\textbf{| Dynamical classification and masses of $z\gtrsim2$ galaxies. } \textit{Left panel:} The observed kinematic ratio of the projected velocity range of a galaxy over the velocity dispersion of the system ($\Delta v_{\rm obs}/2\sigma_{\rm tot}$) as a function of stellar mass. Our measurements at redshift $z=6.8$ are indicated in red points and the SINS sample of H$\alpha$ emitting galaxies at $z\sim2$\cite{ForsterSchreiber2009} is shown in blue squares. Galaxies with $\Delta v_{\rm obs}/2\sigma_{\rm tot}>0.4$ are classified as likely rotation-dominated systems, while sources with $\Delta v_{\rm obs}/2\sigma_{\rm tot}<0.4$ are likely dispersion-dominated (grey line)\cite{ForsterSchreiber2009}. If the velocity gradient observed in these galaxies is due to rotation, these sources are expected to be similar in their dynamical properties to the turbulent, rotation-dominated disks seen for massive galaxies at redshift $z\sim2$. \textit{Right panel:} The comparison of the dynamical or total mass within a $\sim2\,\rm kpc$ half-light radius, when assuming a circularly-symmetric thin disk model, and the stellar mass of our sources (red points). The stellar-mass fractions (grey dotted lines)  of 14 and 43\% for COS-3018555981 and COS-2987030247, respectively, are in good agreement with the range of values found for galaxies in the AMAZE survey\cite{Gnerucci2011} at $z\sim3$ (blue squares) and the SINS survey\cite{ForsterSchreiber2009} at  $z\sim2$ (grey points).
} 
\label{fig:rotdisp}
\end{figure*}
    
The [\ion{C}{2}]\,$\lambda$158$\micron$ line is an important coolant for the neutral interstellar gas. For local galaxies the [\ion{C}{2}] line luminosity is well correlated with the star-formation rate of galaxies\cite{deLooze2014}. In Fig. \ref{fig:SFRCII} we present the measured flux of the [\ion{C}{2}] lines as a function of SFR, which is consistent with the $z\sim0$ SFR-$L_{\rm [CII]}$ relation\cite{deLooze2014}, and consistent with similarly bright galaxies observed at $z\sim5-6$\cite{Capak2015,Willott2015}. In contrast, [\ion{C}{2}] observations in the Epoch of Reionization to-date have shown these galaxies are significantly below the local relation\cite{Ouchi2013,Ota2014,Maiolino2015,Pentericci2016,Knudsen2016}. This is likely due to the distinctly different selection of our targets compared to previous studies at $z>6.5$, where we select [\ion{O}{3}]$\lambda\lambda$5007,4959{\AA}+H$\beta$ emitters as opposed to Ly$\alpha$ emitting galaxies.

Our sources have slightly higher star-formation rates and redder UV slopes ($\beta_{\rm UV}\sim-1.2$)  than previously studied galaxies at this epoch, which could indicate more evolved and higher metallicity galaxies. Sources with extremely low oxygen abundance in the local Universe are typically found to be  [\ion{C}{2}]-deficient \cite{deLooze2014,Smith2017} due to their hard radiation field and therefore metallicity could be an important discriminator between [\ion{C}{2}]-bright and [\ion{C}{2}]-faint sources\cite{Vallini2017}. 
 Moreover, in local galaxies the star formation rate density drives a continuous trend of deepening [\ion{C}{2}] deficit as a function of increasing $\Sigma_{\rm SFR}$\cite{Smith2017,DiazSantos2017}, indicating local processes such as the radiation field intensity are important in driving [\ion{C}{2}] luminosity. If [\ion{C}{2}]-faint sources at $z>6$,  currently unresolved in [\ion{C}{2}], have higher star-formation surface brightness than our galaxies, this could also explain the different SFR/$L_{\rm [CII]}$ ratios.

Furthermore, our sources have  inferred high equivalent-width optical emission lines, which could suggest an ongoing starburst and potentially a high fraction of [\ion{C}{2}] emission emerging from  \ion{H}{2} regions. Starbursts and \ion{H}{2} galaxies in the local Universe have slightly elevated [\ion{C}{2}] luminosities for a given SFR\cite{deLooze2014} and therefore we could specifically be targeting the brightest [\ion{C}{2}] galaxies of the overall $z\sim7$ galaxy population. Finally, while we do not have  spectroscopy covering the Ly$\alpha$ line on COS-3018555981 and COS-2987030247, our sources could be weaker Ly$\alpha$ emitters than typically seen in spectroscopically confirmed sources at this redshift. Ly$\alpha$ emission is suggested to be inversely correlated with neutral gas column density\cite{Stark2017} and can therefore affect the visibility of [\ion{C}{2}], which emerges both in the diffuse neutral and the warm ionized medium of a galaxy.

We determine the [\ion{C}{2}] half-light radii (deconvolved from the beamsize) that extend  $2.6\pm0.8$ and $3.1\pm1.0$ kpc  for COS-3018555981 and COS-2987030247, nearly twice the size of the UV in the brightest LBGs at this redshift\cite{Bowler2017}. We use the spatial extent of the [\ion{C}{2}] detection to investigate the velocity structure of these sources, which reveals a projected velocity difference over the galaxy of  $111\pm28$ and $54\pm20$ $\rm km\,s^{-1}$ for COS-3018555981 and COS-2987030247 respectively (Fig. \ref{fig:kin}), similar to the velocity gradients observed in two galaxies recently studied at $z\sim5-6$\cite{Riechers2014,Pavesi2016}. 
Given the low angular resolution of the observations, there are various ways to interpret these velocity gradients. A rotating galaxy disk would be one interpretation of these velocity fields, however, a merger of one or more [\ion{C}{2}] emitting galaxies, smoothed by the beamsize, could also appear as a regular rotational field. Furthermore, a bipolar outflow, or perhaps an inflow of gas could  provide  an additional velocity component to the [\ion{C}{2}]  line that might give the impression of galaxy rotation. 
 
 We apply an observational criterion for the classification of rotation- and dispersion dominated systems based on the full observed velocity gradient $\Delta v_{\rm obs}$ and the integrated line width $\sigma_{\rm tot}$ of a galaxy, such that $\Delta v_{\rm obs}/2\sigma_{\rm tot}>0.4$ are likely to be rotation-dominated sources\cite{ForsterSchreiber2009}. We compare this observed quantity in our galaxies with those measured from H$\alpha$ in galaxies at $z\sim1-3$\cite{ForsterSchreiber2009} in Fig. \ref{fig:rotdisp}. Despite our sources being an order of magnitude smaller in stellar mass and at an epoch 2.5 billion years earlier in cosmic time we find $\Delta v_{\rm obs}/2\sigma_{\rm tot}$  values of $0.57\pm0.16$ and $0.52\pm0.21$,  similar to the turbulent, yet rotationally supported galaxy disks at $z\sim2$ \cite{ForsterSchreiber2009}. 
 Assuming a circularly symmetric galaxy disk model, we estimate a dynamical mass of $M_{\rm dyn}$ of  $1.0_{-0.2}^{+0.3}$ and $0.4_{-0.3}^{+0.9}\times 10^{10} \, M_\odot$ for COS-3018555981 and COS-2987030247 respectively. Note, however, that the influence of turbulence in these sources could increase the dynamical mass estimates (by at most a factor of 2$\times$). These sources are a factor of $\sim$4--10$\times$ lower mass than the bright, UV-selected sources recently observed at $z\sim5-6$, just $\sim$200--300 Myr later in cosmic time\cite{Capak2015}, which appear otherwise similar in their [\ion{C}{2}] and IR properties (Fig. \ref{fig:SFRCII}).   Furthermore, the stellar mass in our sources makes up $\sim$14 and 43\% of the total dynamical mass that we measure (Fig. \ref{fig:rotdisp}), in good agreement with the 33\% estimated in the UV-selected sources at $z\sim5-6$\cite{Capak2015} and consistent with the wide range of values observed for star-forming galaxies at $z\sim1-3$\cite{ForsterSchreiber2009,Gnerucci2011}. These results indicate a significant gas fraction in the inner few kpc of our galaxies, consistent with hydrodynamical simulations of star-forming galaxies at this epoch\cite{Fiacconi2017}.

In conclusion, we present the first ALMA spectroscopic confirmations of normal star-forming galaxies in the Epoch of Reionization and a measurement of velocity structure using the [\ion{C}{2}]\,$\lambda$157.74$\micron$ emission line. 
  These observations will serve at a pathfinder study that will  enable larger samples of $z\sim7$ galaxies to spectrosopically confirmed with ALMA, while the kinematic mapping of $z>6$ galaxies will adds a remarkable new
 dimension to the study of galaxies in their formative
years.


\hrulefill

\putbib[sample]
\end{bibunit}

\begin{bibunit}[naturemag]
\setcounter{mybibitem}{30}

\begin{addendum}
 \item This paper makes use of the following ALMA data: ADS/JAO.ALMA\#2015.1.01111.S. ALMA is a partnership of ESO (UK), NSF (USA) and NINS (Japan), together with NRC (Canada) and NSC and ASIAA (Taiwan) and KASI (Republic of Korea), in cooperation with the Republic of Chile. The Joint ALMA Observatory is operated by ESO, AUI/NRAO and NAOJ. This work is part of a Rubicon program "A multi-wavelength view of the first galaxies" with project number 680-50-1518, which is financed by the Netherlands Organisation for Scientific Research (NWO). R.M. and S.C. acknowledge ERC Advanced Grant 695671 ``QUENCH'' and
support by the Science and Technology Facilities Council (STFC).

 \item[Author Contributions] RS performed the data reduction and nearly all data analysis for this work. SC carried out the modelling of a thin disk to the velocity fields. RS led the telescope proposal to obtain the data set, with a number of key ideas contributed by RB. RS is responsible for making all the figures and has written the majority of the text of this manuscript. All  co-authors have contributed feedback on the various versions of this manuscript. 
 
 \item[Author Information] The authors declare that they have no competing financial interests. Correspondence and requests for materials
should be addressed to Renske Smit~(email: rs940@cam.ac.uk).

 \item[Code and data availability statement]
The data used in this study was reduced and partly analysed with the public code CASA, available at https://casa.nrao.edu/casa\_obtaining.shtml. The reduction pipeline for this source can be downloaded as part of the ALMA observations with project code 2015.1.01111.S, available at the archive at https://almascience.nrao.edu/alma-data/archive.  The kinematic models used for this study are available from the corresponding author upon request. 
\end{addendum}

\vspace{1cm}

\centerline{\textbf{\sc Methods}}




\subsection{Definitions.}
Throughout this letter we adopt a Chabrier\cite{Chabrier2003} initial mass function (IMF). 
For ease of comparison with previous studies we take $H_0=70\,\rm km\,s^{-1}\,Mpc^{-1}$, $\Omega_{\rm{m}}=0.3$, and $\Omega_\Lambda=0.7$, which gives a physical scale of 5.3\,kpc/{"} at $z=6.8$. Magnitudes are quoted in the AB system\cite{OkeGun1983}.

\subsection{Data.}
We obtained ALMA observations centred on the sources COS-3018555981 (R.A. = 10:00:30.185, decl. = +02:15:59.81) and COS-2987030247 (R.A. = 10:00:29.870, decl. = +02:13:02.47) as part of a filler program (project code: 2015.1.01111.S, PI: Smit) on April 14, 2016, in Cycle-3. Three tunings were requested to cover the frequency range 1870.74-1971.43 GHz in band 6, in order to scan for [\ion{C}{2}] at redshift $z=6.45-6.90$, corresponding to the 99\% photometric redshift probability range\cite{Smit2015}. One tuning was executed, scanning the redshift range $z_{\rm [CII]}=6.74-6.90$, with 24 min on source integration time for each of the targets. The precipitable water vapor (PWV) of the observations were 1.34 mm. The array consisted of 36 antennas and three spectral windows having a bandwidth of 1.875 GHz were used to cover a frequency range of 4.95 GHz in a single sideband. 

The data were calibrated and reduced with the Common Astronomy Software Application (CASA)\cite{McMullin2007} version 4.5.3,  using the automated pipeline, and we imaged the data with the CLEAN task (no iterations as no continuum sources are detected in the data), using a natural weighting for optimal signal-to-noise. The resulting observations reached an image rms sensitivity of 0.32 mJy beam$^{-1}$ at 243 GHz in a 50 km/s channel in both pointings. The primary beam has a resolution of 1.1"$\times$0.7"  (PA = $-$48$^{\circ}$) for both targets.

We also make use of the $HST$ imaging in the WFC3/F160W ($H_{160}$) and the photometry of these objects that was used in the selection of these galaxies in previous work\cite{Smit2015}. 

\subsection{Line detections.}
\textit{COS-3018555981.} We extract a spectrum from the ALMA cube centred on the rest-frame UV continuum of the galaxy detected in the $HST$  $H_{160}$ band of COS-3018555981 as a first guess and find a clear line detected at $\sim$242 GHz, removed from any atmospheric absorption features and with a peak flux $>3.5\sigma$ above the local noise. Next, we extract spectrally averaged map between 241.85 and 242.10 GHz, which reveals that the emission line is centred on a faint wing of the UV continuum detection, 0.27\arcsec\, removed from the brightest UV clump (Fig. \ref{fig:COS3}). This offset is similar to the typical uncertainty in the $HST$ astrometry of 0.2\arcsec\cite{Dunlop2017}, however, if instead the offset is real this could quite reasonably suggest that the brightest star-forming region in the UV does not spatially coincide with the dynamical centre of the system. 

We determine the significance of the detection by measuring the flux on the spectral line-averaged map in a 1.1\arcsec$\times$0.7\arcsec\, aperture corresponding to the full width at half max of the beam and we repeat this measurement 9000 times at randomly selected positions of the image, resulting in an estimated signal-to-noise ratio of 8.2. To determine the redshift of COS-3018555981 we extract a new one-dimensional spectrum from all pixels above the half maximum of the line detection on the spectral line-averaged map and we fit a Gaussian to the observed line to determine a line centre of $241.97\pm0.01\,\rm GHz$, corresponding to a [\ion{C}{2}] redshift of $6.8540\pm0.0003$, and a linewidth of 232$\pm$30 $\rm km\,s^{-1}$ FWHM (Fig. \ref{fig:COS3}). The only lines other than [\ion{C}{2}]\,$\lambda$158$\micron$ that are expected to be bright enough to be able to explain our detection are  [\ion{O}{1}]\,$\lambda$63$\micron$ and [\ion{O}{3}]\,$\lambda$88$\micron$. However, the  [\ion{O}{1}]\,$\lambda$63$\micron$ and [\ion{O}{3}]\,$\lambda$88$\micron$ redshifts of $18.6$ and $13.02$, respectively, are inconsistent with the $HST$ photometry of this source\cite{Smit2015}. Furthermore, the photometric redshift of $6.76\pm0.07$\cite{Smit2015} is also inconsistent with the [\ion{O}{1}]\,$\lambda$145$\micron$ redshift of 7.5, which is the closest infrared line in frequency, if many times fainter, to [\ion{C}{2}]\,$\lambda$158$\micron$.   

\textit{COS-2987030247.} Similar to the procedure for COS-3018555981 we first search for an emission line in the spectrum extracted over the rest-frame UV continuum of COS-2987030247. We find a tentative narrow line at 243.4 GHz, 40 MHz removed from an atmospheric absorption feature at 243.5 GHz, where the rms is elevated by 1.5$\times$ with respect to the median rms in the data-cube. The spectral line-averaged map extracted between 243.35 and 243.45 GHz shows a $>5\sigma$ detection close to the position of the $HST$ counterpart, i.e. the peak of the map is 0.17\arcsec\, removed from the UV-continuum emission (Fig. \ref{fig:COS3}).  

By sampling the noise in the spectral line-averaged map in ellipsoidal apertures of the beamsize, we measure a signal-to-noise ratio of 5.1 for the detected line at 243.5 GHz, suggesting that the line is indeed a real detection. To further test the significance of the line we perform a blind line search of the data-cube. For each pixel in the cube we extract a one-dimensional spectrum from averaging all pixels within the ellipsoidal aperture of the beamsize and we fit any tentative lines in the spectrum with a Gaussian. If the difference between the $\chi^2$ of the line fit and that of a straight line is greater than 25 (i.e. 5$\sigma$) we extract a velocity-averaged image over the FWHM of the line and inspect the significance of the detection on this image. To remove spurious line detections we again assess the significance of any potential line from the random sampling of the flux in ellipsoidal apertures on the line map. While we robustly detect the line over COS-2987030247,  we find no other sources with a $>5\sigma$ detection in both the one-dimensional spectrum and the spectral line-averaged map. This test, in combination with the small spatial offset from our $HST$ target, confirms that our line-detection over COS-2987030247 is real, and not due to a spurious detection showing up close to the rms peak of the atmospheric absorption feature. 

We extract a new spectrum from all pixels with a flux above the half-max flux in the spectral line-averaged map and use this to measure a spectroscopic redshift of $z_{\rm [CII]}=6.8076\pm0.0002$ for COS-2987030247, in good agreement with the photometric redshift of $z_{\rm phot}= 6.66\pm0.14$.

\begin{figure*}[h]
\centering

\includegraphics[width=0.95\textwidth,trim=0mm 0mm 0mm 0mm] {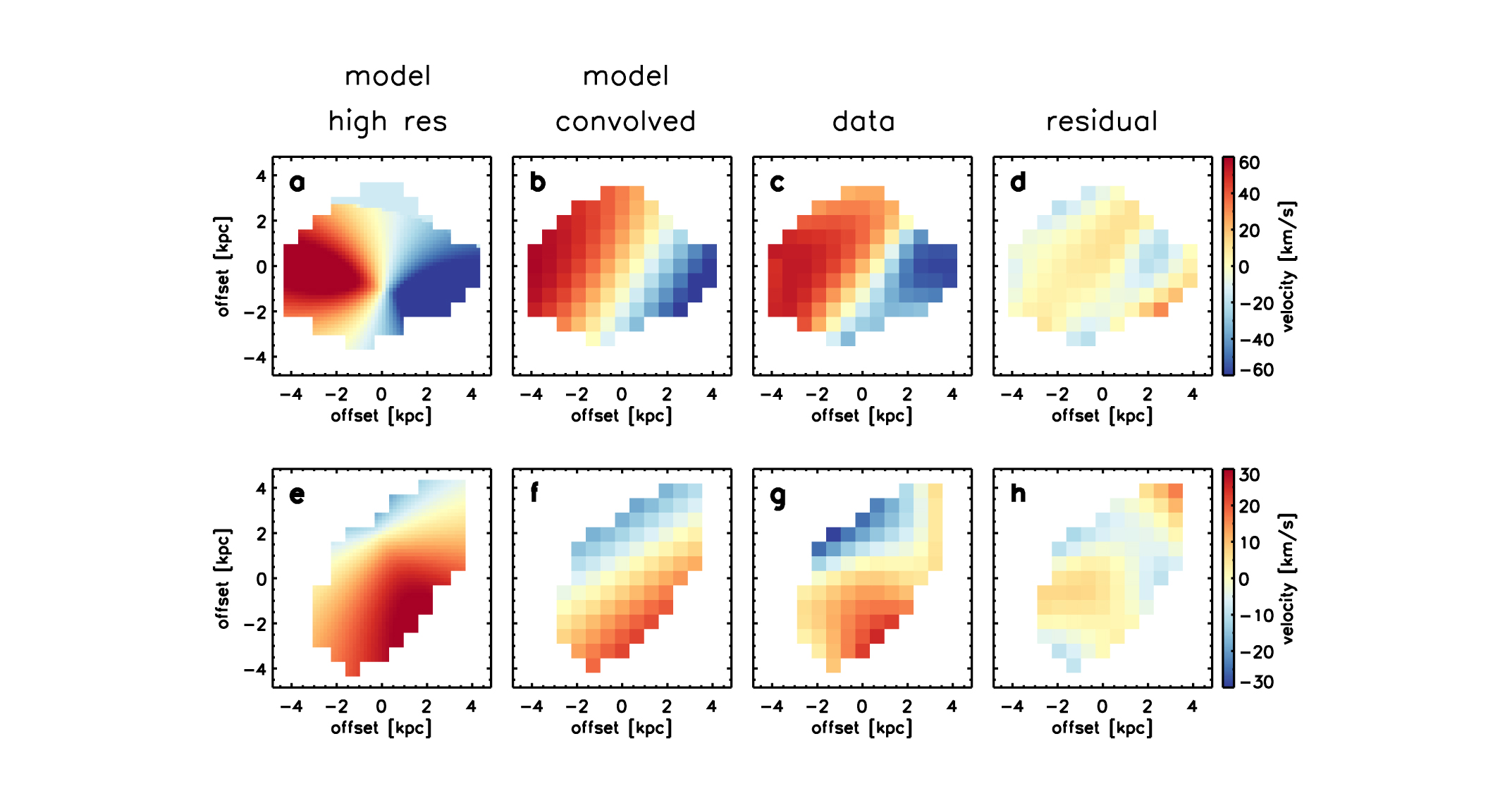} 
\caption{\textbf{| Best fit disk model to the velocity field of the galaxies in this study.} The model fits to the velocity gradients in COS-3018555981 (\textit{top panels}) and  COS-2987030247 (\textit{bottom panels}), when assuming the gas is rotating in an exponential circularly-symmetric thin disk. From left to right the panels show the high-resolution disk model before convolution with the beam, the disk model at the resolution of our observations, our velocity maps as shown in Fig. \ref{fig:kin} and the residuals after subtraction of the model. While the disk model is not a unique solution for these velocity fields, our galaxies are well described by regular rotation. 
}
\label{fig:model}
\end{figure*}

\subsection{Dust.}
We obtain dust continuum measurements after identification of the [\ion{C}{2}] line in our data, by averaging the remaining part of the data-cubes in frequency. We do not find any evidence for flux above the 1$\sigma$ noise-level in the mean continuum image at the source positions. Therefore, we put an upper limit on the continuum flux and assuming a grey body approximation for the dust continuum by considering a range of IR slopes where we vary both the slope in the range $\beta_{\rm IR}=1-2$ and the dust temperatures in the range $T_{\rm dust}=20-60\,\rm K$. We derive a 3$\sigma$ upper limit on the IR luminosity of 1.3 and 1.1$\times10^{11}\,L_\odot$ for COS-3018555981 and COS-2987030247 respectively. 

Since the UV-continuum of galaxies is significantly attenuated by even small amounts of dust, the comparison of the UV colour and the infrared excess, IRX=$L_{\rm UV}/L_{\rm IR}$, can provide insights into the dust attenuation curve in these young galaxies. We derive the UV-continuum slope $\beta_{\rm UV}$, where $f_\lambda\propto\lambda^{\beta_{\rm UV}}$, from a power-law fit to the $HST$ $J_{125}$ and $H_{160}$ photometry and find values of $-1.22\pm0.51$ and $-1.18\pm0.53$ for COS-3018555981 and COS-2987030247 respectively. Often, the interpretation of the infrared excess as a means to constrain the dust attenuation curve can be affected by the geometry of the dust\cite{Narayanan2017}. In particular, a spatial offset between dust-obscured star-forming regions and unobscured UV emitting regions can produce bluer UV colours for a given IRX\cite{Koprowski2016}. The small spatial offsets measured between the UV continuum and [\ion{C}{2}] emission in our sources, might indicate such a dust geometry in this study. However, given that our sources already appear significantly redder than would be predicted  by the Meurer\cite{Meurer1999} relation for a given IRX, our conclusions are not affected by any spatial offsets of the dust continuum with respect to the UV light.

\subsection{Star formation rate and Stellar mass.}
We obtain constraints on the UV star-formation rates from the $J_{125}$ band photometry, corresponding to rest-frame $\sim$1600{\AA}, and on the IR star-formation rates from the upper limits on the IR luminosity, and we convert from luminosity to star-formation rates using the Kennicutt\cite{Kennicutt2012} scaling relations. For COS-3018555981 a $z=0.74$ foreground object is visible at a projected distance of 2.6\arcsec which could introduce a small boost to the measured fluxes due to gravitational  lensing. However, the stellar mass of this object is only $4\cdot10^9\,M_\odot$\cite{Brammer2012}, which suggests a modest halo mass and therefore we estimate the magnification of this source to be no more than 0.1 magnitude (i.e. no larger than the measured random errors) as recently discussed in the literature\cite{Bowler2017}. 

Using the deconvolved size of the [\ion{C}{2}] emission as the size of the galaxy we find a star-formation rate density, $\Sigma_{\rm SFR}$, of 0.91 and $0.75 \,M_\odot \rm\,yr^{-1}\,kpc^{-2}$. This is in good agreement with the star-formation rates obtained using [\ion{C}{2}] as a spatially resolved star-formation rate indicator using the relation calibrated for galaxies from the local KINGFISH sample \cite{HerreraCamus2015}, which predicts a $\Sigma_{\rm SFR}$ of 0.68 and $0.34 \,M_\odot \rm\,yr^{-1}\,kpc^{-2}$ based on the [\ion{C}{2}] surface brightness, $\Sigma_{\rm [CII]}$, of 8.5 and $4.6 \cdot 10^{40} \rm \, erg \,s^{-1}\,kpc^{-2}$.

While the rest-frame optical photometry of $z>4$ galaxies can be heavily affected by strong nebular emission lines\cite{Stark2013}, the redshift range $z\sim6.6-7.0$ provides a unique window where the 4.5$\micron$ Spitzer/IRAC band is free of nebular emission line contamination\cite{Smit2014,Smit2015}, providing a good constraint to model the stellar population of galaxies at these redshifts. We use the Bayesian code MAGPHYS\cite{daCunha2008} with the HIGHZ extension\cite{daCunha2015} to fit the stellar population. We include the continuum constraints at 243 GHz, but we remove the 3.6$\micron$ Spitzer/IRAC photometry as this band is affected by high equivalent width nebular emission (EW$_{\rm [OIII]+H\beta}\sim1000-1500${\AA}\cite{Smit2015}). We find that both galaxies have best fit stellar masses around $M_\ast\sim1-2\times 10^9 \,M_\odot$.

\subsection{Velocity structure and dynamical mass.}
The line maps extracted in (Fig. \ref{fig:COS3}) suggest that the [\ion{C}{2}] emission is spatially resolved in both galaxies, which allows us to investigate the presence of any velocity structure in these galaxies.  For the central 4{\arcsec} of the data-cube we extract a one-dimensional spectrum at every pixel, by averaging all the flux within an elliptical aperture the size of the beam centred on the pixel. We fit a Gaussian to these spectra, using the parameters from the fit to the integrated spectrum as initial parameters. We require the fit to the one-dimensional spectrum to be significant at $>5\sigma$. 

We measure a projected velocity difference over the galaxy of $111\pm28$ and $54\pm20$ $\rm km\,s^{-1}$  for COS-3018555981 and COS-2987030247  respectively, from the minimum and maximum central frequencies taken from the fits that are significant at $>5\sigma$. Galaxies with $\Delta v_{\rm obs}/2\sigma_{\rm tot}>0.4$ (using the measured line widths in Table \ref{tab:lines} to estimate the integrated velocity dispersion) can be classified as likely rotation dominated systems in cases where the data quality prevents reliable kinematic modelling\cite{ForsterSchreiber2009}. This is an approximate diagnostic based on simulations of disk galaxies with a wide range of intrinsic properties. The observed limit of $\Delta v_{\rm obs}/2\sigma_{\rm tot}\sim0.4$ corresponds to the intrinsic ratio of $v_{\rm rot}/\sigma_0=1$\cite{ForsterSchreiber2009}.  We test the robustness of the observed velocity gradient by re-imaging the ALMA data with CASA, using a Briggs weighting with a robustness parameter 0.5, which produces images of the [\ion{C}{2}] emission at lower signal-to-noise, but slightly improved spatial resolution (0.9\arcsec$\times$0.7\arcsec). We confirm that the same analysis on the higher resolution data still produces a velocity gradient with the same projected velocity difference over the two galaxies.   

We will assume that these galaxies can be described by symmetric rotating disks. This is a reasonable assumption given the consistent prediction of high-resolution hydrodynamical zoom simulations that cool gas indeed settles into regular rotating disks\cite{Feng2015,Fiacconi2017,Katz2017,Pallottini2017} and the prevalence of disks among star-forming galaxies at lower redshifts \cite{Stott2016,Wuyts2016,Mason2017}. To derive a dynamical mass for these systems, we adopt two methods. First, we use 
use the approximation that the dynamical mass is estimated from  $M_{\rm dyn}(r<r_{1/2})=(v_{\rm d}^2r_{1/2})/G$, where $v_{\rm d}$ is derived from the averaged of the observed velocity gradient over the galaxy $v_{\rm d}\sin(i)=1.3\Delta v_{\rm obs}$ and the integrated velocity dispersion $v_{\rm d}\sin(i)=0.99\sigma_{\rm tot}$\cite{ForsterSchreiber2009}. We estimate a half-light radius and the inclination of the system from an ellipsoidal fit to the [\ion{C}{2}] emission line map using CASA (corrected for the beam) and find $r_{1/2}$ of  $2.6\pm0.8$ and $3.1\pm1.0$ kpc and $\sin(i)$ of  $0.59\pm0.15$ and $0.88\pm0.06$ for our sources. We derive dynamical masses of  $25.3\pm15.4\times10^{9}\, M_\odot$ and $3.4\pm1.7\times10^{9}\, M_\odot$ for COS-3018555981 and COS-2987030247  respectively.

For a second mass estimate, we model the velocity field assuming that the gas is rotating in a circularly symmetric thin disk, with a gravitational potential that depends only on the disk mass and assuming an exponential surface mass density distribution. The circular velocity is projected along the line of sight, weighted by the intrinsic line surface brightness profile and convolved with the beam size of the observations. Free parameters of our model are the inclination of the disk, the position angle of the disk line of nodes, the systemic velocity of the galaxy and the dynamical mass, measured in a radius of 5\,kpc.  Our method has been successfully applied to ALMA observations of [\ion{C}{2}] emitting sources at $z\sim5$ \cite{Carniani2013,Williams2014}.  Our free parameters are simultaneously constrained from the velocity maps using least-squared fitting. Furthermore, we fit the coordinates of the disk centre based on the surface brightness maps, which is a minor uncertainty on our final results. We estimate uncertainties from the $\chi^2$ parameter space, which is constrained with Monte Carlo Markov chain simulations. The best fit model describes our velocity field well, leaving small residuals, see Fig. \ref{fig:model}. The best fit parameters indicate half-light radii of 1.7$_{-0.3}^{+0.4}\,\rm kpc$ and 2.1$_{-1.1}^{+2.1}\,\rm kpc$, inclination angles of  $\sin(i)=0.87_{-0.10}^{+0.07}$ and $\sin(i)=0.64_{-0.30}^{+0.22}$ 
and dynamical masses of  $1.0_{-0.2}^{+0.3}$ and $0.4_{-0.3}^{+0.9}\times 10^{10} \, M_\odot$ for COS-3018555981 and COS-2987030247  respectively. These values are all consistent within the uncertainties with our estimates derived in the previous section. We therefore adopt this more sophisticated method for our fiducial dynamical mass estimates.  

In the methods described above the effect of turbulence on the estimated dynamical masses is not included\cite{Burkert2010,Burkert2016}. For dispersion dominated galaxies the dynamical mass including pressure support can be estimated by $M_{\rm dyn}=2R_{1/2}(v_{\rm rot}^2+\sigma_0^2)/G$\cite{Newman2013}, where $v_{\rm rot}$ is the inclination corrected velocity gradient and we estimate $\sigma_0$ of 55 and 30 $\rm km \, s^{-1}$. The resulting dynamical masses are 0.3 and 0.4 dex higher than our previous estimates for COS-3018555981 and COS-2987030247  respectively. To study the effect of asymmetric drift on the rotation curve in more detail, higher resolution observations will be required.








\begin{table*}
\begin{center}
\begin{threeparttable}
\centering
\caption{Galaxy properties}
\begin{tabular}{lcc} 
\hline 
\hline 
ID & COS-3018555981 & COS-2987030247 \\
$z_{\rm phot}$ & 6.76$\pm$0.07 & 6.66$\pm$0.14  \\
$z_{\rm [CII] }^b$  & 6.8540$\pm$0.0003 & 6.8076$\pm$0.0002 \\
S/N$^a$ & 8.2 & 5.1\\
$[$\ion{C}{2}] line flux $\rm (Jy\,km\,s^{-1})^b$ & 0.39$\pm$0.05 & 0.31$\pm$0.04\\
FWHM$_{\rm [CII]}\,\rm (km\,s^{-1})^b$ & 232$\pm$30  & 124$\pm$18\\
158$\mu$m continuum flux ($\mu$Jy) & $<$87$^c$ & $<$75$^c$ \\ 
$L_{\rm [CII]}$ $(10^{8}\,L_\odot)$  & 4.7$\pm$0.5 & 3.6$\pm$0.5 \\
$L_{\rm UV}$ $(10^{11}\,L_\odot)$  & 1.1$\pm$0.1 & 1.3$\pm$0.1 \\
$L_{\rm IR}$ $(10^{11}\,L_\odot)$ & $<$1.3$^c$  & $<$1.1$^c$  \\
SFR$_{\rm IR}$ $(M_\odot\,\rm yr^{-1})$ & $<$19$^c$  & $<$16$^c$  \\
SFR$_{\rm UV}$ $(M_\odot\,\rm yr^{-1})$ & 19.2$\pm$1.6  & 22.7$\pm$2.0\\
$M_\ast$ $(10^9 \,M_\odot)$ & $1.4_{-0.2}^{+0.7} $ & $1.7_{-0.2}^{+0.5}$ \\
$M_{\rm dyn}$ $(10^{9} \, M_\odot)$ & $10_{-2}^{+3}$ & $4_{-3}^{+9}$ \\
$\Delta v_{\rm obs}/2\sigma_{\rm tot}$ &  $0.57\pm0.16$ & $0.52\pm0.21$ \\
$r_{1/2,\rm [CII]}$ (kpc) &  $2.6\pm0.8$ & $3.1\pm1.0$ \\
$\beta_{\rm UV}$  & $-1.22\pm0.51$ & $-1.18\pm0.53$ \\
{EW([\ion{O}{3}]+H$\beta$) ($\,$\AA)}\cite{Smit2015}  & $ 1424 \pm 143$ & $ 1128 \pm 166$ \\
\hline
\end{tabular} 
\label{tab:lines} 
$^a$ the S/N measured in a beam-sized aperture (centred on the $HST$ counterpart) on a velocity-averaged image extracted over the detected line. 
$^b$ measured from a Gaussian fit to the integrated spectrum within the half-peak-power contour.
$^c$ 3-$\sigma$ limit.
\end{threeparttable}
\end{center}
\end{table*}


\putbib[sample]
\end{bibunit}
\end{document}